\documentclass[11pt,a4paper]{revtex4}

\usepackage{amsthm}
\usepackage{bbm}
\usepackage{amssymb}
\usepackage{amsmath}

\newtheorem{thm}{Theorem}[section]
\newtheorem*{thm*}{Theorem}

\theoremstyle{definition}

\newcommand{\lra}{\longrightarrow}
\newcommand{\Ra}{\Rightarrow}

\newcommand{\C}{\mathbb{C}}

\newcommand{\1}{\mathbbm{1}}

\renewcommand{\>}{\rangle}
\newcommand{\<}{\langle}

\DeclareMathOperator{\trace}{tr} 
\DeclareMathOperator{\range}{range}
\DeclareMathOperator{\maximize}{maximize}

\begin{document}

\title[Contractivity of positive and trace preserving
maps]{Contractivity of positive and trace preserving maps under $L_p$ norms}
\author{David P{\'e}rez-Garc{{\'\i}}a}
\affiliation{Max Planck Institut f{\"u}r Quantenoptik,
Hans-Kopfermann-Str. 1, Garching, D-85748,
Germany}\affiliation{\'{A}rea de Matem{\'a}tica Aplicada,
Universidad Rey Juan Carlos, C/ Tulipan s/n, $28933$ M{\'o}stoles
(Madrid), Spain}
\author{Michael M. Wolf}\affiliation{Max Planck Institut f{\"u}r Quantenoptik,
Hans-Kopfermann-Str. 1, Garching, D-85748, Germany }
\author{Denes Petz}\affiliation{Alfr\'ed R\'enyi Institute of Mathematics, Hungarian Academy of Sciences
POB 127, H-1364 Budapest, Hungary}
\author{Mary Beth Ruskai} \affiliation{Department of Mathematics,  Tufts University, Medford MA 02155, USA}

\begin{abstract}
We provide a complete picture of contractivity of trace preserving
positive maps with respect to $p$-norms. We show that for $p>1$
contractivity holds in general if and only if the map is unital.
When the domain is restricted to the traceless subspace of
Hermitian matrices, then contractivity is shown to hold in the
case of qubits for arbitrary $p\geq 1 $ and in the case of qutrits
if and only if $p=1,\infty$. In all non-contractive cases  best
possible bounds on the $p$-norms are derived.
\end{abstract}

\maketitle

\section{Introduction}
This paper deals with the following question:\vspace{3pt}
\begin{center}
\parbox{10cm}{\emph{ Given a positive and trace preserving linear map
$T$ between matrix spaces, when is $T$ contractive with respect to the $p$-norm, with
$1\le p\le \infty$ ?}\vspace{3pt}}
\end{center}

This problem has come up in several contexts in recent years. For instance, Olkiewicz
\cite{Olkie}, in his investigation of the superselection structure of dynamical
semigroups, needs as a starting point the fact that a $2$-positive map that is
contractive with respect to both the trace and operator norm  is also contractive with
respect to the Hilbert-Schmidt norm. The same result is needed by Raginsky in
\cite{Rag} in the study of entropy production of a quantum channel. In the context of
quantum information this question arose again in \cite{Vedral}, in the study of
entanglement measures. It is shown there that any distance (in the space of matrices)
that is contractive under completely positive trace preserving maps gives rise to  a
``suitable" entanglement measure. Their conjecture that the Hilbert-Schmidt norm is
such a distance was disproved soon later by Ozawa in \cite{Oz}. In \cite{Nielsen},
Nielsen stated (without proof) that the Hilbert-Schmidt distance is contractive in the
space of qubits, with respect to any completely positive trace preserving map. He also
encouraged further study of this problem.
Recently, the fact that a completely positive trace preserving map is contractive with
respect to the trace norm was
  used in \cite{VeCi05} in the context of condensed
matter theory in a theoretical justification for the high accuracy of renormalization
group algorithms.

Motivated by the appearance of the above question in so many different areas of
physics, we will try in this note to give a complete picture of the solution. We will
first study the general case and then restrict the domain of the maps to the traceless
hyperplane.

\section{The general case}

In the following $\mathcal{M}_n$ will denote the space of $n\times n$ matrices.
  A linear map  $T:\mathcal{M}_n\lra \mathcal{M}_r$
   is called \emph{positive} if it maps positive semi-definite matrices
to positive semi-definite matrices, {\em trace-preserving} if $\trace T(A) = \trace A$
for all $A \in \mathcal{M}_n$, and {\em unital} if $T(\1) = \1$.
   It is easy to see that $T$ is trace-preserving if
   and only if its  adjoint $T^* :\mathcal{M}_r\lra \mathcal{M}_n $ is
unital,
   and that $T$ is positive if and only if $T^*$ is positive.

The $p$-norm (we will assume always $1\le p\le \infty$) of a
matrix $A$ is defined as $\big(\trace |A|^p \big)^{1/p} =
\left(\sum_i \lambda_i ^p\right)^{\frac{1}{p}}$, where the
$\lambda_i$ are the singular values of $A$ (i.e., the eigenvalues
of $|A| \equiv \sqrt{A^*A}$). We write $\mathcal{S}_p^n$ for
$\mathcal{M}_n$ endowed with the $p$-norm. For
$T:\mathcal{M}_n\lra \mathcal{M}_r$, we use $\|T\|_{p-p}$ to
denote the operator norm of $T$ when we consider the $p$-norm in
both the original and the final space, i.e., $\displaystyle{
\|T\|_{p-p}=\sup_{A\in\mathcal{M}_n} \frac{
\|T(A)\|_p}{\|A\|_p}}$. $T$ is called \emph{contractive} under the
$p$-norm if $\|T\|_{p-p}\le 1$.
   Our first result is

   \begin{thm}  \label{thm1}
If $T:\mathcal{M}_n\lra \mathcal{M}_r$ is positive and trace preserving, then
$\|T\|_{p-p}\le n^{1-\frac{1}{p}}$.
\end{thm}
  Moreover,   the bound $n^{1-\frac{1}{p}}$ is attained when
$T$ is the the trace operator $\trace:\mathcal{M}_n\lra \C$ (which is completely
positive and trace preserving).

The main ingredient  in the proof is a non-commutative version of the Riesz-Thorin
Theorem. (See \cite{Bergh-Lof} or Section IX.4 of \cite{Reed-Simon}.) We will also use
a theorem of Russo and Dye  \cite[Corollary 2.9]{Pau}.
\begin{thm}[Non-commutative Riesz-Thorin]\label{R-T}
If $T:\mathcal{M}_n\lra \mathcal{M}_r$ is a linear map, then
$$\|T\|_{p-p}\le
\|T\|_{1-1}^{\frac{1}{p}}\|T\|_{\infty-\infty}^{1-\frac{1}{p}}.$$
\end{thm}
\begin{thm}[Russo-Dye] \label{RD}
If $T:\mathcal{M}_n\lra \mathcal{M}_r$ is positive, then
$\|T\|_{\infty-\infty}=\|T(\1)\|_{\infty}$.
\end{thm}

\begin{proof} To prove Theorem~\ref{thm1}, first note that under
its hypotheses, $T^*$ is positive and unital. Then Theorem~\ref{RD} implies that
$\|T^*\|_{\infty-\infty}=\|T^*(\1)\|_{\infty} = \| \1 \|_{\infty} = 1$. Hence, using
the duality $(S_1^n)^*=S_{\infty}^n$, we can conclude that $\|T\|_{1-1}=1$.  Moreover,
$\|T\|_{\infty-\infty}=\|T(\1)\|_{\infty}\le \|T(\1)\|_1= \|\1\|_1=n$.   Combining
these bounds with Theorem \ref{R-T} gives the result claimed result.
\end{proof}
  We used the fact that when $T$ is trace preserving, then $T$
positive implies  $\|T\|_{1-1}=1$.   In  \cite[Proposition 2.11]{Pau} it is shown that
for $T$ trace preserving,   $T$  is positive if and only if  $\|T\|_{1-1}=1$.

When $T$ is positive, trace preserving and unital  the argument used to prove
Theorem~\ref{thm1} shows that $\|T\|_{1-1}=  \|T\|_{\infty-\infty}=1$.  Then
Theorem~\ref{R-T} implies that $T$ is contractive for all $p$-norms. The next Theorem
shows that this is an equivalence.
\begin{thm}\label{Thmunital}
If $T:\mathcal{M}_n\lra \mathcal{M}_n$ is positive and trace preserving, the following
are equivalent:
\begin{enumerate}
\item[i)] $\|T(\1)\|_p\le n^\frac{1}{p} $ for some
$1<p\le \infty$.

\item[ii)] $T$ is unital.

\item[iii)] $T$ is contractive for the $p$-norm for every $1\le
p\le \infty$.

\item[iv)] $T$ is contractive for the $p$-norm for some $1<p\le
\infty$.
\end{enumerate}
\end{thm}

\begin{proof}
It only remains to prove that  (i) $\Ra$ (ii).   To do this, let $(\lambda_i)_{i=1}^n$
denote the eigenvalues of $T(\1)$. Since $T$ is positive, $\lambda_i\geq 0$; and since
$T$ is trace-preserving, $\sum_i \lambda_i= \trace T(\1) = \trace \1 = n$. H\"older's
inequality can then be used to conclude that
  $\sum_i\lambda_i^p\geq n$
with equality if and only if $\lambda_i=1$ for all $i$.
  But, by assumption,
$\|T(\1)\|_p\le n^\frac{1}{p}$ for some $p>1$.  Thus,
  we  must have equality so that $T(\1)=\1$.
\end{proof}

The hypothesis that $T$ is both unital and trace-preserving can
only be satisfied when $r = n$.   In that case, when $T$ is
trace-preserving, but not unital, it follows that $ \|T\|_{p-p} >
1$.   When $n \neq r$, this does not hold, i.e., there are
non-unital trace-preserving completely positive maps
$T:\mathcal{M}_n\lra \mathcal{M}_r$ for which $ \|T\|_{p-p} < 1$.
To see this one needs Jencova's result  \cite{J}
   that $\|T\|_{p-p} = \omega_p(T^C)$ where $\omega_p(T)$ is
the completely bounded $1 \rightarrow p$ norm studied in
\cite{DJKR} and $T^C$ denotes the conjugate or complementary
channel defined in \cite{Hv} and \cite{KMNR}. From the results in
\cite{DJKR} one can find depolarizing channels $T_{\rm dep}$ such
that $\omega_p(T_{\rm dep}) <1$. To see this let $\mu =
\tfrac{1}{n+1}$ in eq. (5.4) in \cite{DJKR}. Since $\mu =
\tfrac{1}{n+1}$ is the boundary between depolarizing channels
which are entanglement-breaking and those which are not, this
yields examples in both classes. Since the conjugate $T_{\rm
dep}^C$ is {\em not} unital  \cite{KMNR}, we have explicit
examples of non-unital trace-preserving completely positive maps
$T:\mathcal{M}_n\lra \mathcal{M}_{n^2}$ for which $\|T\|_{p-p} <
1$.

The implication   (ii) $\Ra$ (iii) was proved using  complex interpolation.
  For $p =2$, one can obtain an elementary proof by using
  that $\|T\|_{2-2}^2$ is the largest eigenvalue value of
   $T^* \! \circ T$ considered as an
  operator on the Hilbert space $\mathcal{M}_n$ with inner product
  $\langle A, B \rangle = \trace A^* B$.    When $T$ is both
trace-preserving
  and unital, $\1 $ is an eigenvector with eigenvalue $1$, and the
  orthogonality of eigenvectors implies that $\trace G  = \trace \1  G =
0$
  for any other eigenvector  $G$ (which we can assume is Hermitian
without
  loss of generality).
  Now let $G$ be one of these eigenvectors
and let $\omega$ be the largest real number for which $\1 + \omega
G$ is positive semi-definite.   Since $T^* \! \circ T$ is also
positive, $  (T^* \! \circ T)(\1 + \omega G) = \1 + \lambda \omega
G \ge 0$. But this implies that $\lambda \leq 1$ by the definition
of $\omega$ so that $ \| T^* \! \circ T \|_{\infty} = 1$.

\bigskip


\section{The traceless hyperplane}

Using the $p$-norm   to measure the distance between density
matrices, gives expressions of the form $\|\rho-\rho'\|_p$, where
$\rho-\rho'$ is a Hermitian matrix with trace $0$. In this section
we investigate the behavior of such distances under positive and
trace preserving maps.   Let $T|_{\mathcal{H}_0}$ denote the
restriction of $T$   to the hyperplane $\mathcal{H}_0$ of
traceless Hermitian matrices.

\begin{thm}\label{H01}
Let $T:\mathcal{M}_n\lra \mathcal{M}_n$ be a positive trace preserving linear map. Then
  $$\|T|_{\mathcal{H}_0}\|_{p-p}\le \left\{\begin{array}{cc}
    \left(\frac{n}{2}\right)^{1-\frac{1}{p}}, & n \text{\ even} \\
    \left(\frac{2^{2-p}}{(n-1)^{1-p}+(n+1)^{1-p}}\right)^{1/{p}}, & n
\text{\ odd} \\
  \end{array}\right.$$

  Moreover, this bound is optimal, since there exists a completely
positive trace
   preserving map that saturates the inequality.
\end{thm}
\begin{proof}
    We begin by proving the upper bound. For an arbitrary positive trace
preserving map
     $T:\mathcal{M}_n\lra \mathcal{M}_n$, consider $A$ Hermitian,
traceless and with $\|A\|_p\le
     1$. We can write $A= A_+ -A_-$ with $A_+$ and $A_-$ both positive
semi-definite and
     $A_+ A_- = 0$.
     Since $T$ is positive, $[T(A)]_+ \le T(A_+)$ and $[T(A)]_-\le
T(A_-)$. Then,
             \begin{eqnarray*}
     \|T(A)\|_p^p & = &  \trace |T(A)|^p =  \trace \big([T(A)]_+\big)^p
+ \trace \big([T(A)]_-\big)^p \\
          &=&   \|  [T(A)]_+  \|_p^p +   \|  [T(A)]_-  \|_p^p  ~  \leq ~
\|T(A_+)\|_p^p+\|T(A_-)\|_p^p.
         \end{eqnarray*}
     Call $r=\range(A_+)$ and $s=\range(A_-)$ and  denote the
eigenvalues of $A_+$ and
     $A_-$ by $\lambda_1,\ldots,\lambda_r$ and $\mu_1,\ldots, \mu_s$
respectively.
     It follows from Theorem~\ref{Thmunital} that $\|T(A_+)\|_p^p\le
r^{p-1} \|A_+\|_p^p$ and
     $\|T(A_-)\|_p^p\le s^{p-1} \|A_-\|_p^p$. Using Lagrange multipliers
in the problem
     $$
\maximize\left\{ r^{p-1}\sum_{i=1}^r\lambda_i^p +s^{p-1}\sum_{i=1}^s\mu_i^p\right\}
     \quad \text{ restricted to}
         $$
         \begin{align*}
&\sum_{i=1}^r\lambda_i^p +\sum_{i=1}^s\mu_i^p=1\\
&\sum_{i=1}^r\lambda_i-\sum_{i=1}^s\mu_i=0
     \end{align*}
     one finds that at least one of the following two conditions is
satisfied:
     $\lambda_i=\lambda_j$ and $\mu_i=\mu_j$ for every $i,j$, or $s=r$.

     In the first case we have that (assuming now w.l.o.g. that
$\trace(A_+)=1$)
         $$\frac{\|T(A)\|_p^p}{\|A\|_p^p}\le \frac{2}{r^{1-p}+
s^{1-p}}\;.$$ This is in turn maximized and leads to the inequality in
Theorem~\ref{H01} if $s=n-r$ and $r=n/2$ for even $n$, and $r=(n+1)/2$ for odd $n$
respectively. In the second case $r=s$ we have that
     $$\|T(A)\|_p^p\le r^{p-1}\left(\|A_+\|_p^p+\|A_-\|_p^p\right),$$
yielding to the sought inequality for $r=n/2$ (even $n$) whereas $r<n/2$ does not lead
to a new inequality.

To prove optimality of the bound above, consider the completely positive and trace
preserving map $T:\mathcal{M}_n\lra \mathcal{M}_n$ given by
\begin{equation*}
T(A)=|0\rangle\langle 0|\trace[P A] + |1\rangle\langle 1|\trace[(\1-P) A]\,,
\end{equation*} where $P$ is a projector of dimension $d=\trace P$.
If we apply this map to a traceless Hermitian operator of the form
$A=P-\frac{d}{n-d}(\1-P)$ we obtain
\begin{equation*}
\frac{\|T(A)\|_p}{\|A\|_p}=\left(\frac{2 d^p}{d+d^p(n-d)^{1-p}}\right)^{1/p}\;.
\end{equation*}
This  achieves the above bound if $d=n/2$ (d=(n+1)/2) for $n$ even (odd).
\end{proof}

Any trace-preserving map can be written uniquely in the form $T(A)
= N \trace(A) + T_1(A)$ where $T_1(A)$ is a unital
trace-preserving map and $N =\tfrac{1}{d}[T(\1) - \1]$  is
traceless.    If $T_1$ is also positive,  it follows from
Theorem~\ref{Thmunital}, that  $\|T|_{\{\trace=0\}}\|_{p-p} \leq
1$ and we can drop the restriction to Hermitian matrices.
Unfortunately, the results above demonstrate that even when $T$ is
positive and trace-preserving,  $T_1$ need not be positive.

\subsection{Maps on qubits}

When $n=2$, Theorem \ref{H01} implies contractivity in the
traceless subspace of Hermitian matrices, i.e.,
$\|T|_{\mathcal{H}_0}\|_{p-p}=1$. Here, however, there is no need
to restrict to Hermitian matrices.   For qubits, $T$ positive and
trace-preserving implies that the  map $T_1$ above is always
positive.
\begin{thm}
For any positive trace preserving linear map $T:\mathcal{M}_2\lra \mathcal{M}_2$  and
$1\leq p\leq\infty$ we have that
  $$\max_{\trace(A)=0, \|A\|_p=1}\|T(A)\|_{p}\le 1.$$
\end{thm}

\begin{proof}
The theorem is  proved by showing that the $T_1$ defined above is
indeed positive.  Consider
  the action of $T$ on a density operator $\rho=\frac12(\1+{
w}\cdot{ \sigma})$ represented as a vector ${ w}\in\mathbb{R}^3$
on the Bloch sphere. Any trace preserving and positive linear map
acts as $$T(\1+{ w}\cdot{ \sigma})=\1+[{ r}+ R{
w}]\cdot\sigma\;,$$ where ${ r}\in\mathbb{R}^3$ and $R$ is a real
$3\times 3$ matrix. $T$ is positive iff $\|w\|_2\leq1$ implies
$\|r+Rw\|_2\leq 1$. Let $\lambda$ be the largest singular value of
$R$. Then there are unit vectors $u,w\in\mathbb{R}^3$ such that
$Rw=\lambda u$. Since $R(-w)=\lambda(-u)$, one can choose the sign
of $w$ such that $r\cdot u\geq 0$, and thus $1\geq \|r+\lambda
u\|_2\geq\lambda$. This implies that the unital trace preserving
map $T_1(\1+{ w}\cdot{ \sigma}):=\1+[R{ w}]\cdot\sigma$ is indeed
positive, and the result follows from Theorem~\ref{Thmunital}.
\end{proof}

\subsection{The case of qutrits}

Theorem \ref{H01} still implies contractivity in $\mathcal{H}_0$
for the case $n=3$ if $p=1$ or $p=\infty$ (while this fails for
$1<p<\infty$). As in the case of qubits one might expect that the
result for $p=\infty$ also extends to non Hermitian matrices. This
is, however, not the case. A simple counterexample is given by the
map
\begin{equation*}\label{eq.traceless.1}
T(A)=\sum_{i=0}^{1}\<i|A|i\> |0\>\<0| + \<2|A|2\> |1\>\<1|.
\end{equation*}
acting on $A=a_0|0\>\<0|+a_1|1\>\<1|+a_2|2\>\<2|$, where $a_0,a_1,a_2$ are the $3$
complex cubic roots of unity. In this case we have that $\trace(A)=0$,
$\|A\|_{\infty}=1$, but $\|T(A)\|_{\infty}>1$.

{\bf Acknowledgement} We thank J. I. Cirac and M. Junge for
valuable discussions, A. Harrow for raising the question of
contractivity on the traceless hyperplane, and A. Jencova for the
remark following Theorem~\ref{Thmunital}.   The work of D.P. and
M.B.R. was partially supported by the U.S. National Science
Foundation under Grant  DMS-0314228. D. P. is supported by the
Hungarian grant OTKA T032662. D. P-G. is supported by Spanish
project MEC MTM-2005. Parts of this work were done when D.P. was
visiting Tufts University, when M.B.R. was visiting the Max-Planck
Institut f\"ur Quantenoptik, and during the workshop on quantum
information in Benasque.

\end{document}